\RequirePackage[left]{lineno}
\documentclass[aps,prd,superscriptaddress,floatfix,nofootinbib,notitlepage]{revtex4}

\usepackage{graphicx}
\usepackage{amsmath}
\usepackage{longtable}
\usepackage{savesym}
\usepackage{aas_macros}
\usepackage{xcolor}
\usepackage{color}
\usepackage{tabularx}

\usepackage[sort&compress]{natbib}
\usepackage{hyperref}

\graphicspath{{./figures/}}

\newcommand{\gammaRays}{$\gamma$ rays}
\newcommand{\gammaRayHyph}{$\gamma$-ray}

\newcommand{\irf}[1]{\texttt{#1}}

\newcommand{\Fermi}{{\textit{Fermi}}}

\usepackage{color}

\ifdefined\bwfigures
\else
\fi

\newcommand\fdg{\mbox{$~.\!\!^\circ$}} 
\newcommand\dg{\mbox{$~\!\!^\circ$}}

\newcommand{\nc}{\newcommand}

\nc{\beq}{\begin{equation}}
\nc{\eeq}{\end{equation}}
\nc{\barray}{\begin{eqnarray}}
\nc{\earray}{\end{eqnarray}}
\nc{\barrayn}{\begin{eqnarray*}}
\nc{\earrayn}{\end{eqnarray*}}
\nc{\bcenter}{\begin{center}}
\nc{\ecenter}{\end{center}}
\nc{\ket}[1]{| #1 \rangle}
\nc{\bra}[1]{\langle #1 |}
\nc{\0}{\ket{0}}
\nc{\mc}{\mathcal}
\nc{\er}[1]{(\ref{eq:#1})}
\nc{\onehalf}{\frac{1}{2}}
\nc{\partialbar}{\bar{\partial}}
\nc{\psit}{\widetilde{\psi}}
\nc{\Tr}{\mbox{Tr}}
\nc{\hc}{\mbox{H.c.}}
\nc{\ev}{\;\mathrm{eV}}
\nc{\mev}{\;\mathrm{MeV}}
\nc{\gev}{\;\mathrm{GeV}}
\nc{\tev}{\;\mathrm{TeV}}

\def\chii0{\chi_i^0}
\def\chij0{\chi_j^0}

\newcommand{\gsim}{\lower.7ex\hbox{$\;\stackrel{\textstyle>}{\sim}\;$}}
\newcommand{\lsim}{\lower.7ex\hbox{$\;\stackrel{\textstyle<}{\sim}\;$}}
\nc{\ttbar}{t\bar t}

\begin{document}

\title{Search for Gamma-ray Emission from p-wave Dark Matter Annihilation in the Galactic Center}

\date{\today}


\author{C.~Johnson}
\email{arcjohns@ucsc.edu}
\affiliation{Santa Cruz Institute for Particle Physics, Department of Physics and Department of Astronomy and Astrophysics, University of California at Santa Cruz, Santa Cruz, CA 95064, USA}
\author{R.~Caputo}
\email{regina.caputo@nasa.gov}
\affiliation{Center for Research and Exploration in Space Science and Technology (CRESST) and NASA Goddard Space Flight Center, Greenbelt, MD 20771, USA}
\author{C.~Karwin}
\author{S.~Murgia}
\affiliation{Physics Department, University of California at Irvine, Irvine, CA}
\author{S.~Ritz}
\affiliation{Santa Cruz Institute for Particle Physics, Department of Physics and Department of Astronomy and Astrophysics, University of California at Santa Cruz, Santa Cruz, CA 95064, USA}

\collaboration{{\it Fermi}-LAT Collaboration}
\noaffiliation

\author{J.~Shelton}
 \email{sheltonj@illinois.edu}
\affiliation{Department of Physics, University of Illinois at Urbana-Champaign, Urbana, IL 61801, USA}%
\noaffiliation

\begin{abstract}
Indirect searches for dark matter through Standard Model products of its annihilation generally assume a cross-section which is dominated by a term independent of velocity ($\it{s}$-wave annihilation). However, in many DM models an $s$-wave annihilation cross-section is absent or helicity suppressed. To reproduce the correct DM relic density in these models, the leading term in the cross section is proportional to the DM velocity squared ($\it{p}$-wave annihilation). Indirect detection of such $p$-wave DM is difficult because the average velocities of DM in galaxies today are orders of magnitude slower than the DM velocity at the time of decoupling from the primordial thermal plasma, thus suppressing the annihilation cross-section today by some five orders of magnitude relative to its value at freeze out. Thus $p$-wave DM is out of reach of traditional searches for DM annihilations in the Galactic halo. Near the region of influence of a central supermassive black hole, such as Sgr A$^*$, however, DM can form a localized over-density known as a ``spike''.  In such spikes the DM is predicted to be both concentrated in space and accelerated to higher velocities, thereby allowing the $\gamma$-ray signature from its annihilation to potentially be detectable above the background.  We use the $\it{Fermi}$ Large Area Telescope to search for the $\gamma$-ray signature of $\it{p}$-wave annihilating DM from a spike around Sgr A$^*$ in the energy range 10 GeV-600 GeV. Such a signal would appear as a point source and would have a sharp line or box-like spectral features difficult to mimic with standard astrophysical processes, indicating a DM origin.  We find no significant excess of $\gamma$ rays in this range, and we place upper limits on the flux in $\gamma$-ray boxes originating from the Galactic Center. This result, the first of its kind, is interpreted in the context of different models of the DM density near Sgr A$^*$. 
\end{abstract}
\maketitle

\section{INTRODUCTION}\label{sec:intro}
There are strong indications that a significant component of matter in the universe is not described by the Standard Model (SM).
Observational evidence for this new, {\em dark} form of matter comes
from its gravitational influence on visible matter in measurements ranging from the early Universe to the present day
\cite{Zwicky:1933gu,Rubin:1980zd,Olive:2003iq,Ade:2013zuv}. The particle properties of dark matter (DM), however, remain elusive.

One of the most straightforward mechanisms to produce DM in the early universe is thermal freezeout. In this scenario, DM has
interactions with other fields, possibly but not necessarily SM
particles, that ensure DM is part of the thermal radiation bath that
fills the early universe. As the universe cools, the DM annihilation rate drops below the Hubble rate and annihilations freeze out, leaving a thermal
relic abundance of DM. The DM annihilation
cross-section is thus directly related to its cosmic abundance, and yields predictions for the residual DM annihilation rate in galaxy halos today. 
 Generically, the leading
contribution to the thermally-averaged DM annihilation cross-section
$\langle\sigma v\rangle$ will be from velocity-independent $s$-wave processes, so that the present-day annihilation
cross-section is the same as its value during thermal freeze
out~\cite{Steigman:2012nb}.
Such $s$-wave thermal cross-sections generally produce $\gamma$-ray and cosmic-ray signals at interesting (and potentially observable) rates.
The \Fermi Large Area Telescope (\Fermi-LAT), for instance, is capable of probing the $s$-wave thermal cross-section for DM masses up to a few hundred GeV across a variety of annihilation channels \citep{albert_searching_2017}.
The latest generation of cosmic-ray detectors (e.g. AMS-02, PAMELA) is similarly sensitive; an observed excess of high-energy antiprotons can be interpreted as the annihilation signal of DM with a thermal cross-section \citep{cui_possible_2017, hooper_what_2015}.  

In many models, however, symmetries
forbid the $s$-wave contribution to the annihilation cross-section,
and the leading contribution to DM annihilations occurs in the
$p$-wave, $\langle\sigma v\rangle \propto v^2$. For instance, charged scalar DM annihilating to the SM through
an $s$-channel gauge boson has its leading contribution in the
$p$-wave as a consequence of angular momentum conservation \cite{Hagelin:1984wv}. Another example is provided by
fermionic Higgs portal DM \cite{Kim:2006af, Kim:2008pp}; here $CP$ (charge and parity)
conservation enforces the vanishing of the $s$-wave annihilation
cross-section. $CP$ conservation also ensures $p$-wave annihilation cross-sections in a broad and natural class of secluded DM models \cite{Pospelov:2007mp,Tulin:2013teo,Shelton:2015aqa,Evans:2017kti}.
In these models, fermionic DM freezes out via annihilations to light (e.g.~pseudo-Nambu-Goldstone)
bosons $\phi$, $\chi\chi \to \phi\phi$, with $\phi$ subsequently
decaying to the SM. Despite their simplicity, these models present an extraordinarily challenging scenario for detection, leading to the moniker ``nightmare" DM.
DM velocities even in
galaxy clusters today are a tiny fraction of what they were at thermal
freeze out. In the Milky Way, typical DM velocities are $v_{gal}\sim
10^{-3} c$, while at thermal freeze out $v_{fo}\sim 1/3$c. Thus the
annihilation rates for $p$-wave DM in the Galactic halo today are
suppressed by a factor of $\sim 10^{-5}$ relative to the expectation
for $s$-wave DM, making astrophysical detection of $p$-wave DM
annihilations largely out of reach: constraints from light element abundances, Cosmic Microwave Background (CMB) observations, radio data, and
\gammaRayHyph~Galactic diffuse emission are orders of magnitude away
from sensitivity to thermal $p$-wave
annihilations~\cite{Essig:2013goa, Diamanti:2013bia,Depta:2019lbe}. For secluded nightmare models, the lack of detectable signals in conventional indirect detection searches is especially concerning, as
 the coupling between $\phi$ and the SM will generically be
parametrically small, easily placing both collider and direct
detection signals out of reach \cite{Pospelov:2007mp} \footnote{In the limited regions of parameter space where nightmare DM interacts sufficiently strongly with itself to form bound states, the $s$-wave signals from bound state decay can provide an indirect detection signature in the CMB \cite{An:2016kie}.}. Given the dismaying ease with which nightmare models evade all traditional searches for DM, it is of high interest to consider other avenues to discover or constrain $p$-wave DM.

Unique opportunities for detecting $p$-wave DM may be offered by the DM density spikes that can form around supermassive black holes (SMBHs). Depending on the formation history of the black hole (BH) and its astrophysical environment, such spikes can yield
extraordinarily dense concentrations of DM, and thus
bright, localized signals, particularly in models of annihilating DM
\cite{Gondolo:1999ef, Gnedin:2003rj, Regis:2008ij,
 Fields:2014pia, Lacroix:2015lxa}.
Critically, the DM velocity dispersion increases inside the spike, $
v^2 (r) \propto M_{BH}/r$, with $r$ the distance from the BH, to
support the power-law increase in density. In other words,
supermassive black holes act as mild DM accelerators, opening a
window onto the physics of thermal freeze out and thereby potentially
enabling the observation of processes that were active in the early
universe but are otherwise inaccessible in the present day
\cite{Amin:2007ir,Cannoni:2012rv, Arina:2014fna, Shelton:2015aqa}.
DM annihilation (or decay) within SMBH-induced density spikes
would appear as a point source to $\gamma$-ray telescopes, with the main
component of the $\gamma$-ray spectrum arising from DM annihilations.
$p$-wave DM annihilation within such spikes can thus give rise to
potentially observable kinematic features in the $\gamma$-ray energy
spectrum with a point-source-like spatial morphology, even in the
absence of a corresponding signal from the
halo~\cite{Shelton:2015aqa}. This probe of
$p$-wave DM annihilation in the $\gamma$-ray spectrum offers a 
discovery tool for models of DM that could otherwise elude detection
entirely.

With this motivation in mind, the \gammaRayHyph~data set compiled by \Fermi-LAT is of great
interest~\cite{Atwood:2009ez}. The \Fermi-LAT is one of the most
sensitive instruments to DM with weak-scale mass and cross section
annihilating into \gammaRays. Analysis of the {\it Fermi}-LAT \gammaRayHyph~data
can place strong limits on, or discover, DM annihilation with cross
sections near the canonical thermal value into a wide variety of
SM particles. However, most recent searches by the {\it Fermi}-LAT Collaboration, including both
searches for a continuum excess and spectral features, have assumed
$s$-wave annihilation processes~\cite{Ackermann:2013uma,
 Ackermann:2015lka, Ackermann:2015zua}, mainly because the velocity
suppression makes searches for $p$-wave annihilation processes
insensitive to thermal relic DM. 

A handful of authors (e.g. \citep{Zhao:2017dln}) have searched for velocity-dependent DM annihilation in dwarf galaxies, but as far as we are aware, a detailed study of velocity-dependent annihilation at the Galactic Center (GC) has not yet been undertaken.
Meanwhile a growing body of work (for instance \cite{Amin:2007ir,Fields:2014pia,Shapiro:2016ypb}) suggests that the GC with its SMBH Sgr A$^*$ may be the best location to search for $p$-wave DM annihilation.

In the \Fermi-LAT energy spectrum, the spikes would contain sharp spectral features such as \gammaRayHyph~lines \cite{Bergstrom:1988fp} or boxes (a flat distribution of photon flux between two energy endpoints) \cite{Ibarra:2012dw}, allowing such a spike signal to be distinguished from known astrophysical sources\footnote{The sharp spectral features may be subdominant}.
A search for sharp spectral features in a point-like source is distinct from the search for line emission in the Galactic halo performed in Ref.~\cite{Ackermann:2013uma}. 
Basic searches for box-like spectral features at the GC have been performed previously \citep{Ibarra:2012dw}, but generally assume a different phenomenology (i.e. a velocity-independent annihilation cross section, and therefore a different spatial morphology) than what is considered here.

In this paper, we consider the \gammaRayHyph~emission from the GC, which is host to the SMBH Sgr A$^*$.
Specifically, we focus on the core of Sgr A$^*$, where the flux from $p$-wave DM annihilation is expected to come from, and search for both narrow line-like boxes and wide boxes.
Because the $\gamma$-ray spectrum is a falling powerlaw, the sensitivity to wide boxes is driven almost entirely by the sharp feature provided by its upper endpoint.
Therefore, results obtained for a particular wide box may be reliably applied to boxes of intermediate widths as well.

In Section~\ref{sec:dm}, we describe the DM distribution in the GC and
how it relates to searches for indirect signals of DM annihilation.
In Section~\ref{sec:GC}, we discuss the {\it Fermi}-LAT instrument,
the method of modeling the GC as a $\gamma$-ray source, and the data set
and background models used for the DM analysis. The analysis
techniques and the resulting bounds are shown in
Sections~\ref{sec:analysis} and ~\ref{sec:results}, and we conclude in
Section~\ref{sec:conclusion}.

\section{DARK MATTER MODELS}\label{sec:dm}

A black hole (BH) at the center of a DM halo contracts the matter within its zone of influence into a power-law overdensity or ``spike'', $\rho_{DM}(r)\propto r^{-\gamma_{sp}}$ \cite{1972GReGr...3...63P,Gondolo:1999ef,Nakano:1999ta,Ullio:2001fb}.
The steepness of this spike depends on the properties of the DM halo as well as the formation history of the BH, yielding power-laws as shallow as $\gamma_{sp} =1/2$ in the case of BHs that are not at the dynamical center of their surrounding halo \cite{Nakano:1999ta,Ullio:2001fb}, and as steep as  $\gamma_{sp} =2.75$ for BHs growing adiabatically at the center of an NFW-like halo \cite{Gondolo:1999ef}.
Gravitational scattering from baryonic matter can be important in determining the steepness of the final spike if the stellar distribution within the gravitational zone of influence of the BH is sufficiently dense and cuspy, as may be the case for the SMBH at the center of the Milky Way\cite{Merritt:2003qk, Gnedin:2003rj, Merritt:2006mt}.
In this case, the limiting power law for the DM spike is $\gamma_{sp} = 1.5$, attained when the system has reached equilibrium. If the system is still in the process of equilibrating, then non-equilibrium spikes, characterized by intermediate values of $\gamma_{sp}$, are possible.
Meanwhile strong DM self-interactions would lead to yet other intermediate values of $\gamma_{sp}$ \cite{Shapiro:2014oha}. 
For the Milky Way's central SMBH, there are thus a wide range of possible density spikes, depending on the detailed history of the GC and the nature of DM. 

The combination of the high
DM densities and the increased DM velocities within a SMBH-induced density spike
can make thermal $p$-wave DM annihilations observable around
the Milky Way's SMBH across a wide range of assumptions for the DM
distribution in the GC \cite{Shelton:2015aqa}. For
$p-$wave DM, the point-like source from the BH density spike is the
only observable cosmic-ray signal of DM annihilations; there is no
corresponding detectable signal from the halo.

Especially in the absence of a crosscheck from a halo signal, to
ascribe a DM origin to a point-like $\gamma$-ray source in the busy
environment of the GC, it is critical to search for sharp
kinematic features in the energy spectrum such as $\gamma$-ray lines and
boxes. This section will firstly define a general, parametric model of DM spikes in the GC, and secondly
describe a reference model of $p$-wave DM and its $\gamma$-ray signatures.


\subsection{DM distribution in the GC}

We adopt a fiducial model for the DM distribution in the GC following \cite{Fields:2014pia,Shelton:2015aqa}. We take the
halo to be described by a generalized NFW halo, which in the inner
Galaxy takes a power-law form, $\rho(r) = \rho(r_0)
(r_0/r)^{\gamma_c}$. Typical values of the cusp exponent $\gamma_c$
predicted by DM-only simulations are in the range $0.9\lesssim
\gamma_c\lesssim 1.2$ \cite{Diemand:2008in,Navarro:2008kc}. Larger
values of $\gamma_c$ can arise through the adiabatic contraction of
the central halo following the dissipative collapse of baryons into
the disk \cite{Blumenthal:1985qy,Gnedin:2004cx,Gustafsson:2006gr},
though such large values are somewhat disfavored by recent
observations \cite{Iocco:2016itg}. We treat $\gamma_{c}$ as a free
parameter. We take the solar system to lie at $r_{\odot}=8.46$ kpc
from the GC \cite{Do:2013upa}, and the local density of
DM to be $\rho_\odot = 0.3\,\mathrm{GeV}/{\mathrm{cm}^3}$
\cite{Bovy:2012tw}. Here and below, our adoption of specific values
for galactic parameters should be viewed as a fiducial choice, in the
same spirit as the adoption of specific fiducial halo models in more
traditional searches for DM annihilation products in the
halo.

The DM spike begins growing inside the region $r_b \approx 0.2 r_h$ (where $r_h = GM/v_0^2$ is the radius of gravitational influence of the black hole) \cite{Merritt:2003qc,Merritt:2003qk}, and is well-described as a power law, $\rho_{sp}(r) = \rho_{sp}(r_b) (r_b/r)^{\gamma_{sp}} $.
Here G and M are the Newtonian gravitational constant and the black hole mass, respectively.
As discussed above, different formation histories of the SMBH and the inner Galaxy
yield a wide range of possible values for $\gamma_{sp}$, and we here
consider $\gamma_{sp}$ to be a free parameter. The spike grows
following this power law until it becomes dense enough that
annihilations become important over the lifetime of the spike
$\tau\approx 10^{10}$ years, $\rho_{ann} = m_{\chi}/(\langle\sigma
v\rangle \tau)$. Within the corresponding radius, $r_{in}$, annihilations deplete
the spike and limit the spike's growth to a very mild power law,
$\rho_{in}(r) = \rho_{ann}(r_{in}) (r_{in}/r)^{\gamma_{in}} $. The
inner power law is $\gamma_{in} = 1/2$ for $s$-wave annihilations
\cite{Vasiliev:2007vh}. The increasing importance of $p$-wave
annihilations with decreasing radius further flattens the inner power
law relative to the $s$-wave case; we here adopt the numerical result
$\gamma_{sp} = 0.34$ of Ref.~\cite{Shapiro:2016ypb}. Finally,
the inner boundary of the spike is located at $r_{in}=4 G M$
\cite{Sadeghian:2013laa}.

The DM density in the spike and inner halo is thus modeled as
\cite{Fields:2014pia,Shelton:2015aqa, Shapiro:2016ypb},
\begin{eqnarray}
\rho(r) &=& 0, \ \ \ r < 4GM \ \ ({\rm capture \ region}), \\
&=& \frac{\rho_{\rm sp}(r)\rho_{\rm in}(t,r)}{\rho_{\rm sp}(r) + \rho_{\rm in}(t,r)}, \ 4GM < r < r_b \ \ ({\rm spike}), \nonumber \\
&=& \rho_b(r_b/r)^{\gamma_{c}}, \ \ \ \ \ \ r_b < r < r_H\;\;({\rm inner \ halo})\nonumber
\end{eqnarray}
We take $M= 4\times 10^6 M_\odot$ \cite{Genzel:2010zy,Ghez:2008ms} and
adopt as our reference inner halo dispersion $v_0 = 105 \pm 20 {\rm
 ~km~s^{-1}}$~\cite{0004-637X-764-2-184}, which together determine the
radius $r_b$ at which the spike begins to grow. Our fiducial value of the halo dispersion velocity is about 5\% higher than the value found in \cite{0004-637X-764-2-184}; higher halo velocity dispersion leads to a smaller detectable flux \cite{Fields:2014pia}, so this value is slightly conservative.

To support the power-law increase in density, the velocity dispersion
inside the spike must also increase.  We take the velocity
dispersion as isotropic, and model it by matching a piece-wise continuous
approximate solution of the Jeans equation within the spike onto a
constant in the inner halo, giving
\begin{eqnarray}
v^2(r) &=& \frac{GM}{r} \frac{1}{1+\gamma_{in}} 
\left[
1 + \frac{r}{r_{\rm in}}
\left( \frac{\gamma_{in}-\gamma_{sp}}{1+\gamma_{sp}} \right)
\right], \nonumber \\
&& \ \ \ \ \ \ \ \ \ \ \ \ \ \ \ \ 4GM \leq r < r_{\rm in} \ \
({\rm inner \ spike}), \label{eq:velocity}\\
&=& \frac{GM}{r} \frac{1}{1+\gamma_{sp}}, \ \ \ 
r_{\rm in} \leq r < \frac{r_h}{1+\gamma_{sp}} \ \ ({\rm outer \ spike}), 
 \nonumber \\
&=& v_0^2 = {\rm const},  \ \ \frac{r_h}{1+\gamma_{sp}}\leq r 
\ \ ({\rm cusp }). \nonumber
\end{eqnarray} 

The dominant contribution to the emission from DM annihilations within
the spike occurs at $r_{in}\sim 10^{-3}-10^{-5}$ pc for thermal dark
matter. At this radius, the DM velocity is still non-relativistic,
$v\sim 0.1 c$.

The $\gamma$-ray flux per unit energy from (self-conjugate) DM annihilating within the spike is given by
\beq
\frac{d\Phi_\gamma}{dE_\gamma} = \frac{1}{4\pi R_\odot^2} \,\frac{1}{2 m_\chi^2} \,\frac{dN_\gamma}{dE_\gamma} \int_{4GM}^{r_b} 4\pi r^2 dr\, \rho^2(r) \langle \sigma v (r)\rangle,
\eeq
where $dN_\gamma/dE_\gamma$ is the $\gamma$-ray energy spectrum produced in a single annihilation.
As the density profile of the spike $\rho(r)$ itself depends on the DM
annihilation cross-section through $r_{in}$, the $\gamma$-ray flux from
the spike does not depend linearly on the annihilation
cross-section. For a $p$-wave spike, the flux depends on the
annihilation cross-section as
$\Phi_{sp}\propto (\langle\sigma v\rangle)^{(3-\gamma_{sp})/(1+\gamma_{sp})}$
\cite{Shelton:2015aqa}. Thus spike signals depend more weakly on
the annihilation cross-section than do traditional halo searches.
The results are moderately sensitive to other parameters in the model (the black hole mass and halo velocity distribution); for an estimate of the sensitivity, see Figure 2 of \cite{Fields:2014pia}.

\subsection{A reference $p$-wave DM model}

As a reference model of thermal $p$-wave DM, we adopt here a specific realization of
nightmare DM, the ``Hidden Sector Axion Portal (HSAP)'' model of
\cite{Shelton:2015aqa}. In this model DM is a Majorana fermion $\chi$
that annihilates to pairs of pseudo-scalars $\phi$, which subsequently
decay to SM gauge bosons.  The Lagrangian describing the interactions of the DM and the pseudo-scalar is given by
\beq
\mathcal{L}=\bar\chi (i\gamma\cdot\partial)\chi - m_\chi \bar
  \chi \chi  +\frac{1}{2}(\partial \phi)^2  -\frac{1}{2} m_\phi^2 \phi^2 - i y
  \phi\,  \bar \chi \gamma^5 \chi ,
\eeq
where $m_\chi$, $m_\phi$ are the masses of $\chi$, $\phi$, and $y$ is the Yukawa coupling that will govern the annihilation cross-section.
$CP$ conservation in this model ensures that the leading contribution to the DM annihilation cross section occurs in the $p$-wave.
If the spectrum additionally contains a $CP$-even scalar $s$ with $m_S < 2 m_\chi-m_\phi$, then DM annihilation can proceed through the $s$-wave $\chi\chi\to s \phi$ channel
\cite{Nomura:2008ru,Ibarra:2013eda}, but $s$ may easily be too heavy to participate in DM annihilation, or indeed entirely absent. In this case $CP$ forbids the $s$-wave contribution.

\begin{center}
\begin{figure}
\begin{minipage}{6in}
  \centering
  \raisebox{-0.5\height}{\includegraphics[width=0.45\columnwidth]{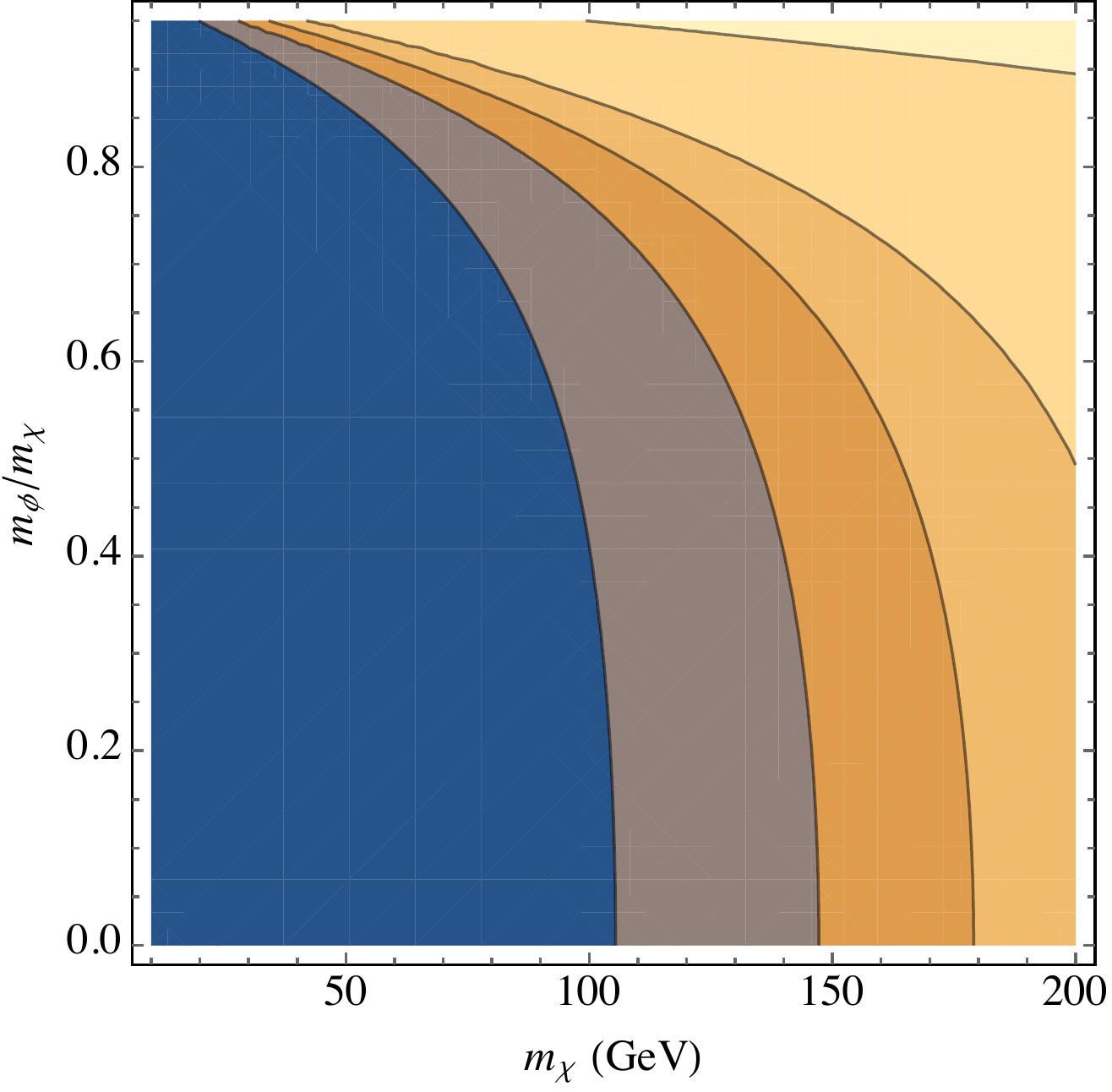}}
  \hspace*{.0in}
  \raisebox{-0.5\height}{\includegraphics[width=0.075\columnwidth]{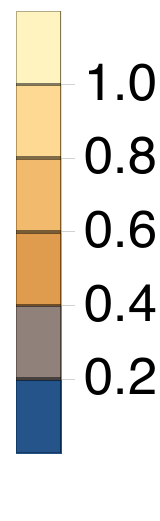}}
\end{minipage}

\caption{Value (represented by the color map) of the coupling $y^4$ required to obtain the
 observed DM abundance in the HSAP model (Eq.~\ref{eq:sigmav}), as a function of $m_\chi$ and
 $\zeta = (m_\phi/m_\chi)^2$. Results in this plot are obtained using an approximate analytic solution to the Boltzmann equation \cite{Kolb:1990vq} and are accurate to $\lesssim 10\%$; for calculations elsewhere in this paper the Boltzmann equation is solved exactly.}
\label{fig:y4}
\end{figure}
\end{center}

 We use this HSAP model to determine the
annihilation cross-section $\langle\sigma v\rangle_{thermal}$ that
yields the observed DM relic abundance as a function of $m_\chi$ and
$\zeta \equiv m^2_\phi/m_\chi^2 $. 
The DM annihilation cross-section is
\beq
\label{eq:sigmav}
\langle \sigma v\rangle = \langle v^2\rangle\,\frac{ y^4}{24\pi m_\chi^2} \sqrt{1-\zeta}
 \,\frac{(1-\zeta)^2}{(2-\zeta)^4} 
\eeq
in the non-relativistic limit
\footnote{For $\zeta \gtrsim 1-\langle v^2\rangle/8$, the velocity dependence in the phase space factor $\sqrt{1-\zeta + \mathcal{O}(v^2)}$ in Eq.~\ref{eq:sigmav} must be retained.
For spikes around the Milky Way's SMBH, $\langle v^2 (r_{in})\rangle \sim 0.01$, and thus the DM annihilation cross-section is still consistently $\propto \langle v^2\rangle$ even for $\zeta=0.99$.
However for $\zeta\gtrsim 0.96$ the velocity dependence in the phase space factor is important for the larger velocity dispersions realized during thermal freezeout, and is retained in our full calculations, where we implement an exact thermal average in a numerical solution of the Boltzmann equation.}. 
The value of $y^4$ needed to obtain the observed DM relic abundance is shown in Fig.~\ref{fig:y4}.

The analogue of the ``smoking gun'' DM line in this nightmare model is a DM {\it box}, i.e., a constant photon flux within
the energy range $\left(1 - \sqrt{1-m_{\phi}^2/m_{\chi}^2},\;1 +
\sqrt{1-m_{\phi}^2/m_{\chi}^2} \right) \times m_{\chi}/2$
\cite{Ibarra:2012dw}. This feature is the result of the decay
$\phi\to \gamma\gamma$, boosted in the Galactic rest frame according
to the kinematics of the annihilation.  Sufficiently narrow boxes appear line-like, while for wide boxes, the upper edge provides a sharp spectral feature that can allow the box to be cleanly identified
above falling continuum backgrounds. As the dominant contribution to the DM
annihilation signal inside the spike comes from regions where the DM
is still highly non-relativistic, kinematic broadening of the box
feature is negligible in comparison to the \Fermi-LAT energy
resolution.  

The branching fraction for the decay
$\phi\to\gamma\gamma$ controls the normalization of the box signature and depends on the couplings of $\phi$ with the SM.
If $\phi$ couples to the SM through axion-like couplings to electroweak gauge bosons,
$\mathcal{L}_{int} = -(1/\Lambda_1) \phi \epsilon_{\mu\nu\rho\sigma}
B^{\mu\nu} B ^ {\rho\sigma}-(1/\Lambda_2) \phi
\epsilon_{\mu\nu\rho\sigma} W^{a\mu\nu} W ^ {a\rho\sigma}$, where $B^{\mu\nu}$ is the hypercharge
field strength and $W^{a\mu\nu}$ is the field strength for the SU(2)$_L$ gauge bosons,
then its
branching ratio to $\gamma \gamma$ (and, if kinematically allowed,
$\gamma Z$) is $\mathcal{O}(1)$, and the $\gamma$-ray box is the leading
signature of DM annihilation. In other models, for instance where $\phi$ decays to
the SM through a mixing with the Higgs, the $\gamma\gamma$ branching
ratio is suppressed, $\sim 10^{-2}-10^{-3}$, and, while the box feature is still
present, the $\gamma$-ray continuum emission arising from other $\phi$
decay modes will typically yield stronger constraints
\cite{Shelton:2015aqa}.
For simplicity, in this work we take the
branching ratio $Br(\phi \to \gamma \gamma) = 1$, i.e., all of the
annihilation flux appears in a $\gamma$-ray box.

Our implementation of the HSAP model should be understood as a convenient reference model in which one may interpret the results of a search for $\gamma$-ray boxes.
As noted above, gauge invariance generally requires $\phi$ to also decay to $Z\gamma$ and $ZZ$ final states when these modes are kinematically accessible, which reduces the $\gamma \gamma$ branching ratio while adding new box and continuum contributions to the $\gamma$-ray energy spectrum \citep{Ullio:1997ke,Perelstein:2006bq,Gustafsson:2007pc,Cline:2012nw,Rajaraman:2012db}. 
We therefore caution the reader to interpret the results carefully above $m_Z$, as we only consider the case $Br(\phi \to \gamma \gamma) = 1$.
Moreover, if $m_\chi> \frac{3}{2} m_\phi$, DM annihilations in the spike are actually dominated by the higher-order $s$-wave process $\chi\chi\to 3 \phi$.
Thus the widest box that is realized by the hidden sector axion portal model, considered literally, is realized for $\zeta=4/9$.
We emphasize that the experimental sensitivity to wide boxes is dominated by the upper end point, and thus limits on a wide box of a given $\zeta$ can reliably be re-interpreted to limit a wide box with different $\zeta$. 

Meanwhile, once the width of the narrow box signal becomes smaller than the experimental resolution, the signal becomes line-like.  While the search presented here is not optimized for line signals, a narrow box search will have sensitivity to $\gamma$-ray lines as well.  Such $\gamma$-ray lines are predicted by models of $p$-wave DM  where DM annihilates directly to SM final states, such as Higgs portal DM \cite{Kim:2006af,Kim:2008pp}.  However in most such models, direct annihilations into diphotons are highly suppressed, and for Higgs portal DM occur in fewer than $\lesssim 10^{-3}$ of events.  For models where the continuum $\gamma$-ray signal dominates to this degree, requiring that DM annihilations within any SMBH density spike not outshine the observed point sources near the GC will typically lead to a more restrictive constraint than a line or box search \cite{Shelton:2015aqa}.

\section{FERMI-LAT OBSERVATIONS OF THE Galactic CENTER}\label{sec:GC}

{\it Fermi}-LAT is an all-sky pair-conversion telescope which has been successfully observing the $\gamma$-ray sky between a few tens of MeV to more than a TeV for ten years.
Incoming $\gamma$ rays pass through an anti-coincidence detector and convert in a tracker to $e^{+}/e^{-}$ pairs. 
Energy is deposited by the $e^{+}/e^{-}$ pairs in a calorimeter.
The charged particle direction is reconstructed using the information in the tracker, and the energy is estimated from depositions in the calorimeter.  
Detailed descriptions of the {\it Fermi}-LAT and its performance can be found in dedicated papers~\cite{Atwood:2009ez,2013arXiv1303.3514A}. 
In the data selection for the present work, {\it Fermi}-LAT has an integrated exposure of approximately $4.5 \times 10^{11}$ cm$^2$s in the direction of Sgr A$^*$.

\subsection{Data Selection}\label{sec:fermi}
For this analysis, we used nine years of {\it Fermi}-LAT data (2008 August 4 to 2018 July 26) selecting \irf{Pass 8} SOURCE-class events 
in the energy range from 6 GeV to 800 GeV, binned in 50 logarithmically-spaced energy bins and 0.04$\fdg$ angular pixelization. 
The energy range was chosen to avoid the well-known \cite[e.g.][]{GC2016} complexities of modeling the GC at energies of a few GeV.
In addition, the {\it Fermi}-LAT point-spread function (PSF) improves by nearly an order of magnitude between 1 GeV and 10 GeV, which improves its sensitivity to a signal that is localized as a point-like source.
Our analysis considers $\gamma$-ray boxes with upper edges above 10 GeV; we include data between 6 and 10 GeV to avoid possible edge effects.

Our region of interest (ROI) was $2^\circ\times2^\circ$ and centered at Sgr A$^*$.
The small ROI was chosen for two reasons: a) our putative DM signal is a point source spatially coincident with Sgr A$^*$, and the {\it Fermi} 95\% containment radius at 10 GeV is less than 1$^\circ$, so our ROI should contain virtually all of the signal, and b) our analysis relies mostly on searching for sharp spectral features, so contamination of unmodeled nearby point sources was not a particular concern.
We found the farthest point source from Sgr A$^*$ in our ROI, 3FHL J1747.2-2959, had negligible correlation with the parameters of the GC source.
In any case, the resulting model showed no indications that our ROI had any appreciable contamination from sources beyond 1$^\circ$ from Sgr A$^*$.

\begin{figure}[ht] 
\begin{center}
\includegraphics[width=0.9\columnwidth]{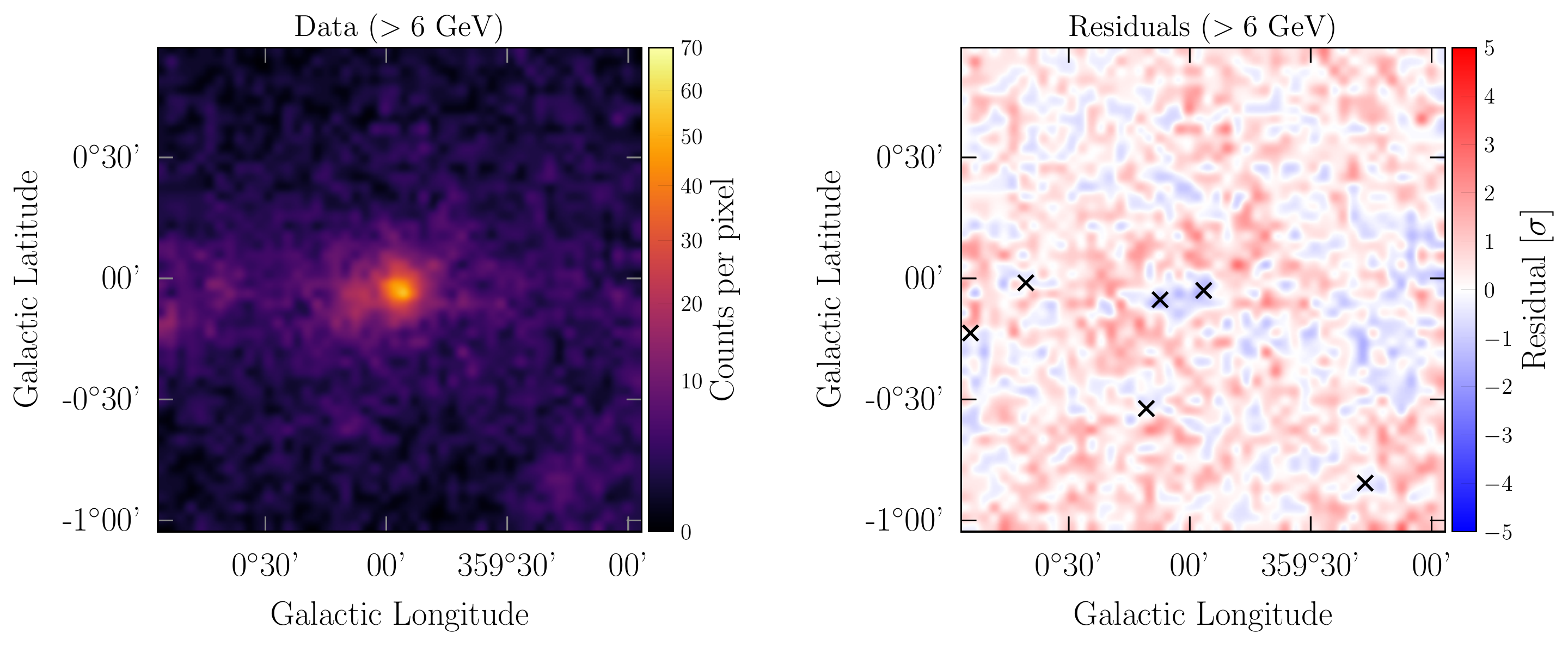}
\noindent
\caption{ 
\label{fig:roi_residmap}
{\it Left panel:} Total photon counts in the ROI used in the analysis. The GC source is prominently seen near the center of the image, while the Galactic diffuse emission is responsible for the majority of the photons outside the GC. {\it Right panel:} Residuals (data-model) in units of $\sigma$ after fitting with {{\tt gtlike}} (see Section \ref{sec:model_GC}). The location of each 3FHL point source in the model is marked with a black (X). No significant excesses or deficits are observed in the data. In both maps, the pixel size is 0.04$\dg$ and a Gaussian smoothing (width of 0.04$\dg$) has been applied.
}
\end{center}
\end{figure}

We modeled the performance of the {\it Fermi}-LAT using the \irf{P8R2\_SOURCE\_V6} Instrument Response Functions (IRFs).
The data processing and exposure calculations were performed using the {\Fermi} Science Tools version 
11r5p3\footnote{\url{http://fermi.gsfc.nasa.gov/ssc/data/analysis/software}}. 
A summary of the parameters of our data selection is available in Table \ref{tab:data2}, and a counts map of the data is shown in the left panel of Figure \ref{fig:roi_residmap}.

\begin{table}[ht]
\begin{tabular}{cc}
\hline \hline
Selection & Criteria \\ \hline
Mission Elapsed Time (s)\footnote{$Fermi$ Mission Elapsed Time is defined as seconds since 2001 January 1, 00:00:00 UTC} & 239557417 to 554321025  \\ 
Instrument Response Functions & P8R2\_SOURCE\_V6 \\
Energy Range (GeV) & 6-800 \\
Fit Region & $2\dg\times2\dg$, centered on (RA, DEC)=(266.417, -29.0079) \\ 
Zenith Range & $\theta_z<$100$\dg$ \\
Data Quality Cut with the $gtmktime$ Science Tool\footnote{Standard data quality selection: \texttt{DATA\_QUAL==1 $\&\&$ LAT\_CONFIG==1}} &  Yes \\ \hline
\end{tabular}
\caption{ \label{tab:data2} Data selection used by this paper's analysis }
\end{table}

\subsection{Modeling the GC}\label{sec:model_GC}

In order to search for a DM signal via the maximum-likelihood analysis described next, in Section \ref{sec:analysis}, we required a model of the ROI.
Our model was built from diffuse components and objects listed in the Third Catalog of Hard \Fermi-LAT Sources (3FHL)~\cite{collaboration_3fhl:_2017}.

\subsubsection{Diffuse Components and Extended Sources}\label{sec:diffuse}
The GC is the most complicated region of the $\gamma$-ray sky, and as a result the parameters of the point source associated with Sgr A$^*$ are dependent on the model of Galactic diffuse emission. 
Although custom interstellar emission models (IEM) have been successfully used to model the GC in past works~\cite{GC2016}, generating a similar custom IEM with the data reconstruction used here was deemed to be outside the scope of this paper, for which we needed only an empirical model against which we can test our DM hypothesis.

The diffuse components used in this analysis were the standard Pass 8 models taken from the Fermi Science Support Center\footnote{The diffuse background models are available at: \url{http://fermi.gsfc.nasa.gov/ssc/data/access/lat/BackgroundModels.html} as \irf{is\_P8R2\_SOURCE\_V6\_v06.txt} and \irf{gll\_iem\_v06.fits}.}.
After an initial fit to the data we found that the contribution by the isotropic component of our model was negligible; we decided not to include an isotropic component in the final model for this reason.
We do not expect its omission to have an impact on the results.
 
The 3FHL catalog comprises sources detected at energies above 10 GeV over the first 7 years of \Fermi-LAT data, and contains 1556 sources.
Six of these sources fall within our ROI, and all have spectra well-described by a power law.
Furthermore, none of the sources in the ROI were found to be extended in the 3FHL catalog.
A summary of all the sources used in the model is shown in Table \ref{tab:model}, and the residuals of the data after optimizing the model are shown in the right panel of Figure \ref{fig:roi_residmap}.
With six point sources and one diffuse component, and two free parameters for each source (the prefactor and spectral index of the power-law), the background-only model contained a total of 14 free parameters.

\begin{table}[ht]
\begin{tabularx}{10cm}{c @{\extracolsep{\fill}} cccc}
\hline \hline
3FHL Source Description  & $N_{\gamma}$ & RA & DEC\\ \hline
Galactic diffuse emission & 4397 & - & - \\ 
J1745.6-2900 & 1253 & 266.42 & -29.01 \\
J1746.2-2852 & 510 & 266.56 & -28.88 \\
J1747.2-2959 & 172 & 266.80 & -30.00 \\
J1747.2-2822 & 137 & 266.82 & -28.37 \\
J1748.1-2903 & 96 & 267.04 & -29.06 \\
J1748.6-2816 & 126 & 267.16 & -28.28 \\
\hline
\end{tabularx}
\caption{ \label{tab:model} List of sources used in modeling the ROI. $N_\gamma$ is the integral number of photons expected from the source, after optimization by {\tt gtlike}.}
\end{table}

As a check of our systematic uncertainty, we also performed the following analysis using a separate dataset and model covering 4 years of data with Pass 7 data reconstruction.
The model of the ROI contained a different set of point sources (from the 3FGL catalog \citep{REF:2015.3FGL}), and diffuse models were taken from the custom IEM of \citep{GC2016}. 
The resulting flux upper limits were found to be consistent with the main analysis presented below; for simplicity we present only our standard analysis.


\section{ANALYSIS}\label{sec:analysis}

\subsection{Fitting Method}\label{sec:fitting}
As discussed in Section \ref{sec:dm}, the phenomenology of our reference model p-wave DM signal is that of a point source located at the location of Sgr A$^*$, with a photon flux that is flat between two endpoints (a `box' shape).
For this analysis, we considered two representative versions of the box: the `wide' box has a value of $\zeta = 0.44$, while the `narrow' box has a value of $\zeta = 0.9999$.
Implications from the two types of searches for mass splittings in intermediate cases are discussed in Section \ref{sec:conclusion}.

We searched for $\gamma$-ray boxes which had an upper-edge energy equal to the boundaries of the energy bins between 10 and 658 GeV in our data selection, corresponding to 42 different DM hypotheses.
In order to prevent potential edge effects from impacting the results, boxes with upper edges outside of this range were not considered. 

The likelihood $\mathcal{L}(n,\theta)$ of a particular model is given by:
\begin{equation}
\label{eq:likelihood}
\mathcal{L}(n, \theta) = \prod_{i=0}^{N}\frac{\mu_i^{n_i}}{n_i!}e^{-\mu_i}
\end{equation}
where the index $i$ runs over the angular and energy bins, and $\mu_i$ and $n_i$ are the predicted and actual photons, repsectively, in bin $i$. 
We varied the model parameters $\theta$ until the likelihood is maximized; in practice we used the logarithm of the likelihood. 
The likelihood computation and maximization was performed by the {\it Fermi} Science Tool \textup{gtlike}, which in turn used the MINUIT \cite{James:310399} optimization routine.

The significance of each DM hypothesis was evaluated using the test statistic (TS) defined as: 
\begin{equation}
\mathrm{TS}=2~\mathrm{ln}\frac{\mathcal{L}(\mu, \theta | \mathcal{D})}{\mathcal{L}_{\mathrm{null}}(\theta | \mathcal{D})} \label{eq:ts}
\end{equation}
Where $\mu$ is the signal strength, $\theta$ is the array of parameters describing the DM hypothesis (in this case, the energy and width of the $\gamma$-ray box, and $\mathcal{D}$ represents the binned data. $\mathcal{L}_{\mathrm{null}}$ is the value of the likelihood in the absence of any signal.
The likelihood values $\mathcal{L}$ are computed from Equation \ref{eq:likelihood}.

The TS value was then used to calculate a level of significance $Z$ via:
\begin{equation}
Z = \Phi ^{-1} \left( 1-\int_\mathrm{TS}^\infty \chi^2(x,k)dx\right)
\end{equation}
Where $\Phi ^{-1}$ is the inverse quantile function; the integral in this expression is the $p$-value.
Simulations (described below) confirmed that the TS values were distributed roughly following a $\chi^2$ distribution with 1 degree of freedom (the total flux contained in the `box' signal){\textemdash}see the left panel of Figure \ref{fig:correlations}. As the number of trials per bin decreases, the $\chi^{2}$ distribution moderately over-predicts the number of high TS trials observed in simulated data. 
An example DM signal with $\zeta = 0.44$ (spatially integrated over the ROI), along with the background, is shown in the left panel of Figure \ref{fig:artificial_box}. 

The procedure for finding the TS of a given DM hypothesis and upper limit on the total flux of a $\gamma$-ray box with an upper edge at a particular bin energy was as follows:
\begin{enumerate}\label{sec:procedure}
\item \label{step:fit} The parameters of the model described in Section \ref{sec:model_GC} allowed to vary to maximize the likelihood function $\mathcal{L}$, giving the null likelihood.
This step was performed once for each dataset under investigation (either the true data or the Monte Carlo simulations described below).
\item The expected spectrum of the DM signal is calculated by convolving a ideal box spectrum with a Gaussian distribution representing with the \Fermi-LAT energy resolution, which is between 5\% and 10\% in the energy range considered. 
\item A point-source with the convolved DM spectrum is added to the model at the location of Sgr A$^*$, with a single overall normalization parameter $N$.
\item All parameters in the model except for the normalization of the central GC source are fixed.
A study of the correlation coefficients (see Section \ref{sec:correlations}) showed that the signal was correlated with this source (especially of DM hypotheses with upper edge below 100 GeV) , but had negligible correlation with other parameters in the model.
Fixing the other parameters also had the benefit of decreasing the computation time and preventing numerical instabilities when fitting a system with a large number of degrees of freedom.
\item The normalization $N$ of the DM source is increased from a value of 0 until the TS exceeds 2.77, which corresponds to the 95\% confidence upper limit on $N$, or a $Z$ value of approximately 2. The value of this TS was computed empirically from the results of the Monte Carlo simulations (see Section \ref{sec:MCMethods}.
This value is approximately the value of the critical $\chi^2$ of 2.71 for a $p$-value of 0.1 with 1 degree of freedom, which is consistent with a one-sided upper limit at 95\% confidence.
The complete likelihood profiles for each DM hypothesis are also stored.
\end{enumerate}

\subsection{Correlations Between Background and Signal Components}\label{sec:correlations}

In order to understand the relationship between a potential signal and the background sources, we calculated the correlation coefficients between the signal source and the GC source.
As expected, both $\zeta=0.9999$ and $\zeta=0.44$ hypotheses are negatively correlated with the normalization of the GC background source.
We found that the signal became less correlated as the right edge of the box increases in energy, since the likelihood fit is strongly driven by the higher statistics at low energy.
We also found that the $\zeta=0.44$ hypothesis had a stronger correlation to the background when compared to the  $\zeta=0.9999$ case, which is expected because the $\zeta=0.44$ signal contributes over a broader energy range. 
A plot of the correlation coefficients in both cases as a function of the energy of the right edge of the box is shown in the right panel of Figure \ref{fig:correlations} below.

We investigated further the degeneracy between the signal and background by recomputing the upper limit on the signal flux with the parameters of all background sources fixed at the value obtained from step \ref{step:fit}.
We cannot say {\it a priori} that the data does not contain any signal, so the solid curves in Figure \ref{fig:brazil_lines} is the main, conservative result.
However, if we were to assume that there was no observed signal, then the dashed curve in Figure \ref{fig:brazil_lines} is the most optimistic limit attainable. 

The prefactor and index describing the power-law spectral shape of the GC source were found to be almost perfectly anticorrelated.
We found that the correlation coefficients of the signal to the parameters of other sources in the model were negligible.

\begin{figure}[ht] 
\begin{center}
\includegraphics[width=0.45\columnwidth]{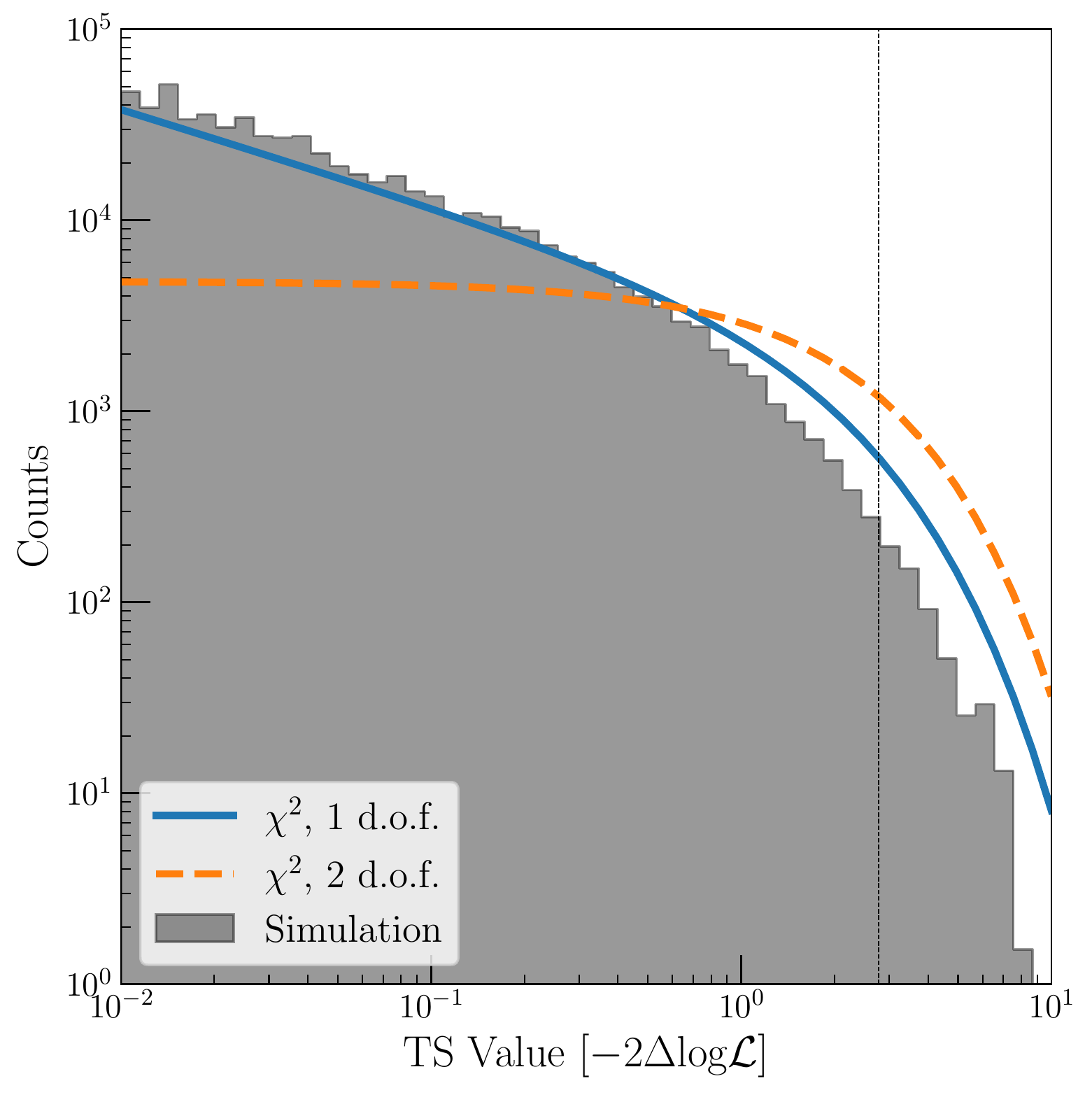}
\includegraphics[width=0.45\columnwidth]{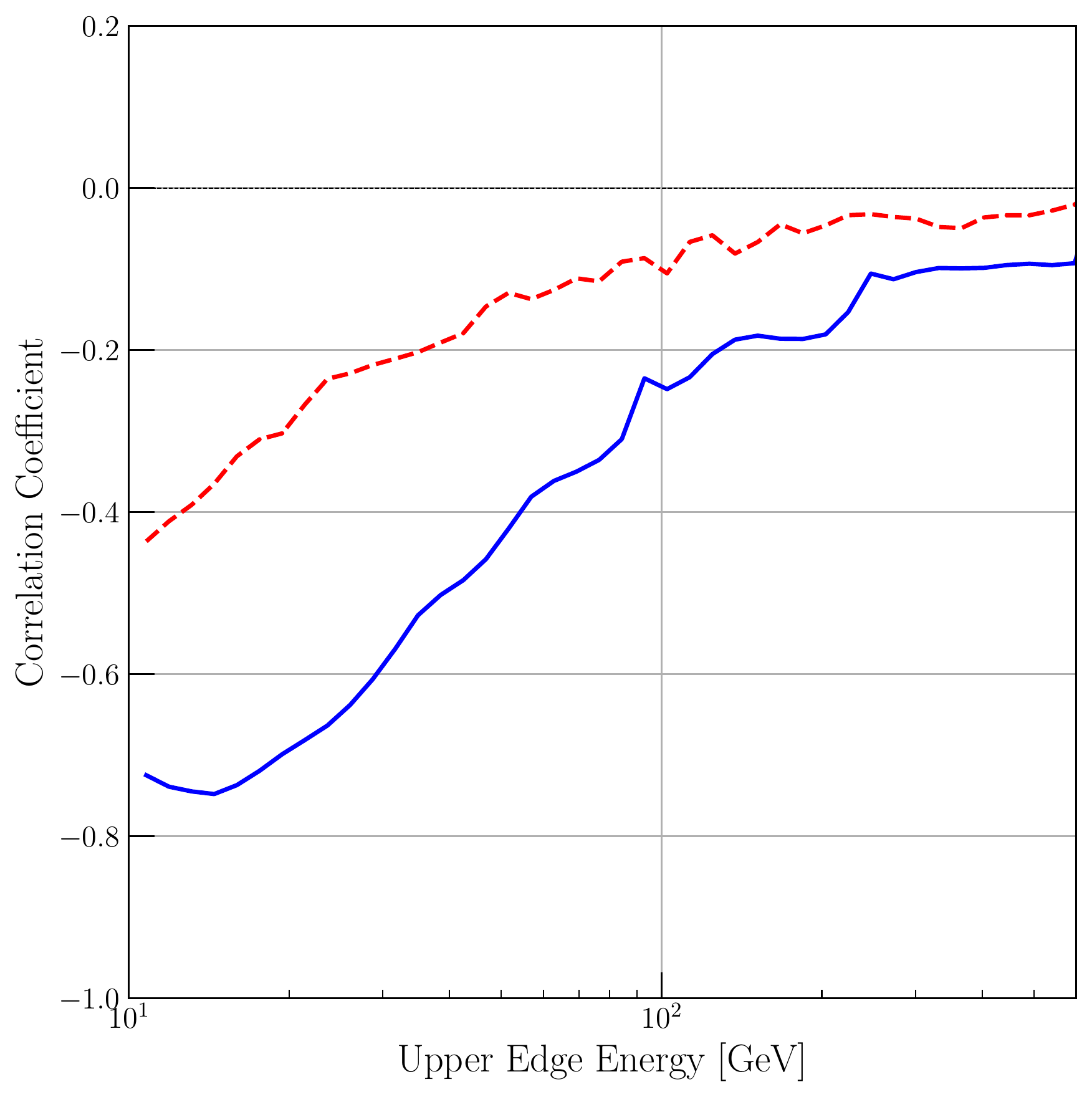}
\noindent

\caption{
{\it Left Panel}: Histogram of TS values for all DM signal hypotheses from the Monte Carlo study for a $\zeta=0.44$ signal. The shape of the TS distribution is well-described by a $\chi^2$ distribution with only one degree of freedom, although the $\chi^2$ distribution slightly overpredicts the Monte Carlo distribution at high TS and underpredicts at low TS.
The critical value of 2.77, is shown as a vertical dashed line.
{\it Right Panel}: Correlation coefficients ($\zeta = 0.44$ in solid blue and $\zeta = 0.9999$ in dashed red) between the total flux of the DM signal hypotheses and the normalization $N$ of the GC source (modeled as a power law, i.e. $\frac{dN}{dE} = Ne^{-\alpha E}$.
We evaluate the correlation as a function of the upper edge of the DM signal box, and find that the correlation is negligible for high-energy boxes but is becomes significant at lower energies because of the increased statistics in the data at lower energies.
Because the two sources are spatially coincident, the sources are expected to be anticorrelated.
}
\label{fig:correlations}
\end{center}
\end{figure}

\subsection{Monte Carlo Simulations\label{sec:MCMethods}}
We performed a Monte Carlo study in order to understand the impact that statistical fluctuations have on the analysis, and to evaluate the distribution of TS values of the signal.
Each instance of the Monte Carlo began by optimizing the background-only model, and generating Poissonian fluctuations around the model.
We then used the Poisson data as the input to the protocol defined in Section \ref{sec:fitting}, and stored the likelihood profiles for each DM hypothesis.
Only MC instances in which the fitting procedure converged with no errors were used in performing the calculations.
Because Step \ref{step:fit} above fits the parameters of the background model, this technique probes the effects of statistical uncertainty on both the signal and the background.

From the sample of MC instances, we found the distribution of TS values that corresponded to the best-fit fluxes of the DM signal. 
The distribution is approximately distributed as a $\chi^2$ with one degree of freedom, which is consistent with the result expected from Wilk's theorem (see the left panel of Figure \ref{fig:correlations}).
The critical TS of 2.77 is shown in the figure as a dashed vertical line.

We performed O($10^3$) simulations, and calculated the upper limit curves from each instance.
The family of curves was used to generate 68\% and 95\% containment bands for the cases of $\zeta = 0.44$ and $\zeta = 0.9999$.
The results are displayed in Figure \ref{fig:brazil_lines}.

\subsection{Reconstruction of Injected Signal\label{sec:injected}}
To confirm that the upper limit calculation was sensitive to the presence of a DM signal, and to understand how a signal would appear in our analysis, we injected a DM signal into the data and repeated the analysis procedure from Section \ref{sec:procedure}.
The injected DM signal for this test was defined to have $\zeta = 0.44$ and a total flux of $1.5\times10^{-10} $ph cm$^{-2} $s$^{-1}$, with an upper energy endpoint of 100 GeV.
At 100 GeV, the ratio of the injected signal flux to the total flux in the ROI was about 30\%.
We performed the same Monte Carlo study on the injected-signal dataset to produce containment bands for the limit.

\begin{figure}[ht]
\begin{center}
\includegraphics[width=0.45\columnwidth]{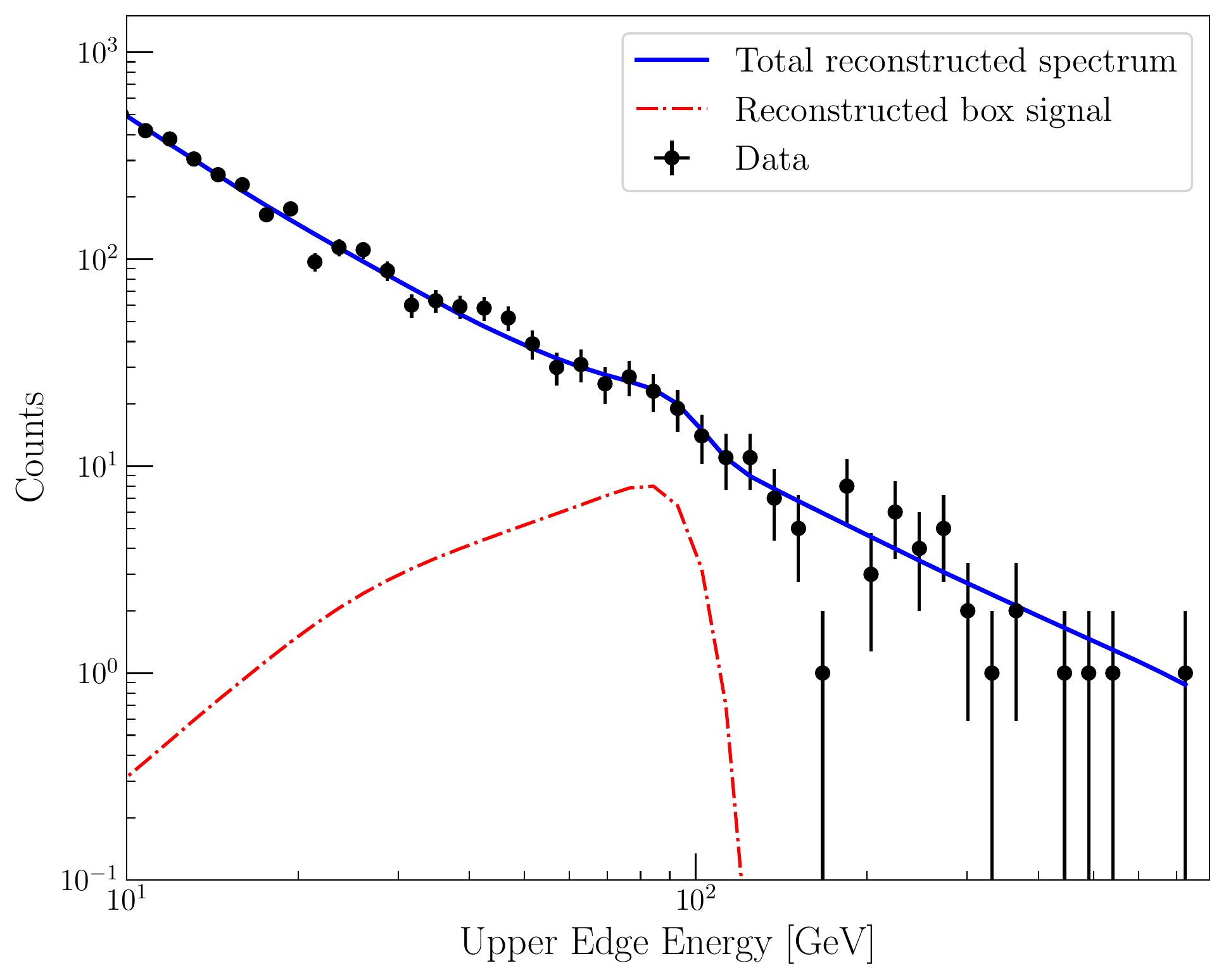}
\includegraphics[width=0.45\columnwidth]{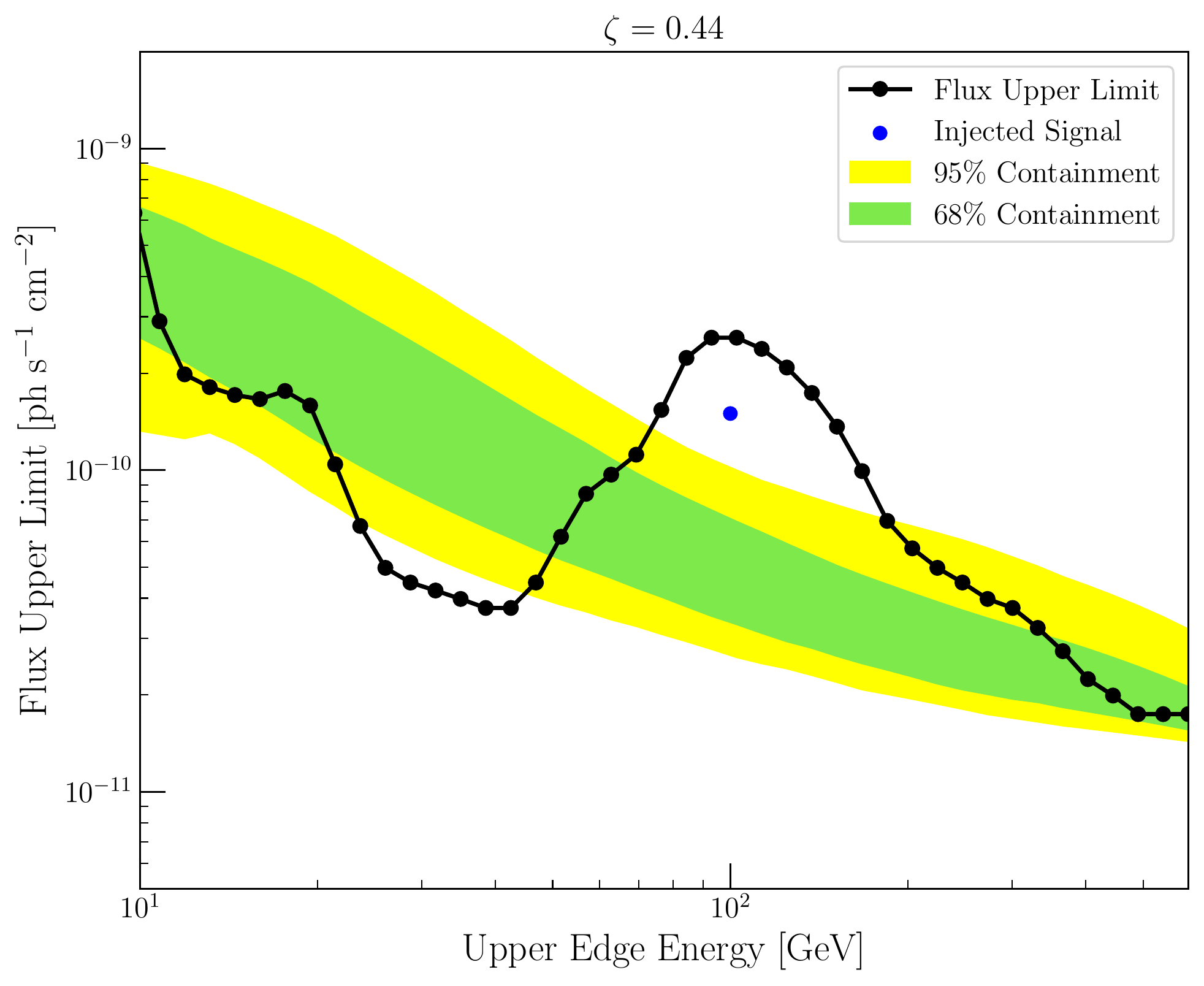}
\noindent
\caption{
\label{fig:artificial_box}
{\it Left Panel}: Energy spectrum of the data + injected signal. The injected `box' DM signal appears in the data as a small bump near its upper endpoint. 
{\it Right Panel}: The DM signal upper limit (in black) in the presence of an injected box with upper endpoint 100 GeV and total flux $1.5\times10^{-11} $ ph cm$^{-2} $s$^{-1}$. The blue dot shows the position of the injected signal.
The 68\% and 95\% containment bands are constructed from performing the analysis on Poisson-fluctuated datasets about the best-fit background model. 
Our injected DM signal is not excluded by the analysis. 
}
\end{center}
\end{figure}

The results of the analysis are in good agreement with the known injected signal.
The best-fit DM hypothesis was found to have an upper edge energy of 102 GeV, and the reconstructed flux of the signal was $1.61\times10^{-10}$ ph cm$^{-2} $s$^{-1}$.
The upper limit curve was found to contain a prominent bump near 100 GeV which noticeably exceeded the 68\% and 95\% containment bands from the Monte Carlo study, as seen in Figure \ref{fig:artificial_box}.
We concluded that the analysis procedure defined in Section \ref{sec:fitting} is sensitive to the presence of a realistic DM signal, and can accurately reconstruct its parameters.
For illustration, the spectrum of a best-fit box with total flux  $3.0\times10^{-10} $ph cm$^{-2} $s$^{-1}$ (double that of the injected box test) is shown in the left panel of Figure \ref{fig:artificial_box}.

\section{Results}\label{sec:results}

\begin{figure}[ht] 
\begin{center}
\includegraphics[width=0.45\columnwidth]{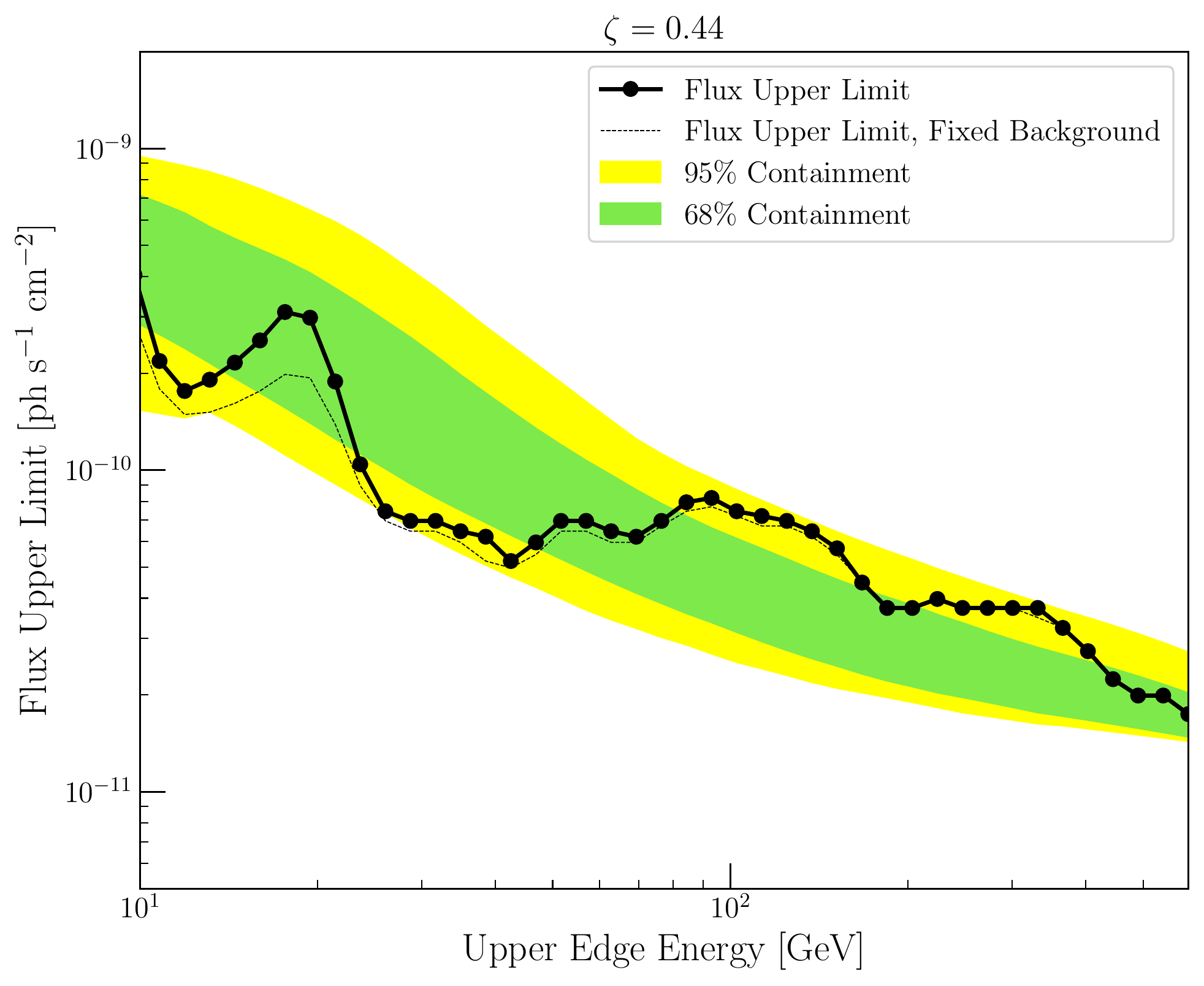}
\includegraphics[width=0.45\columnwidth]{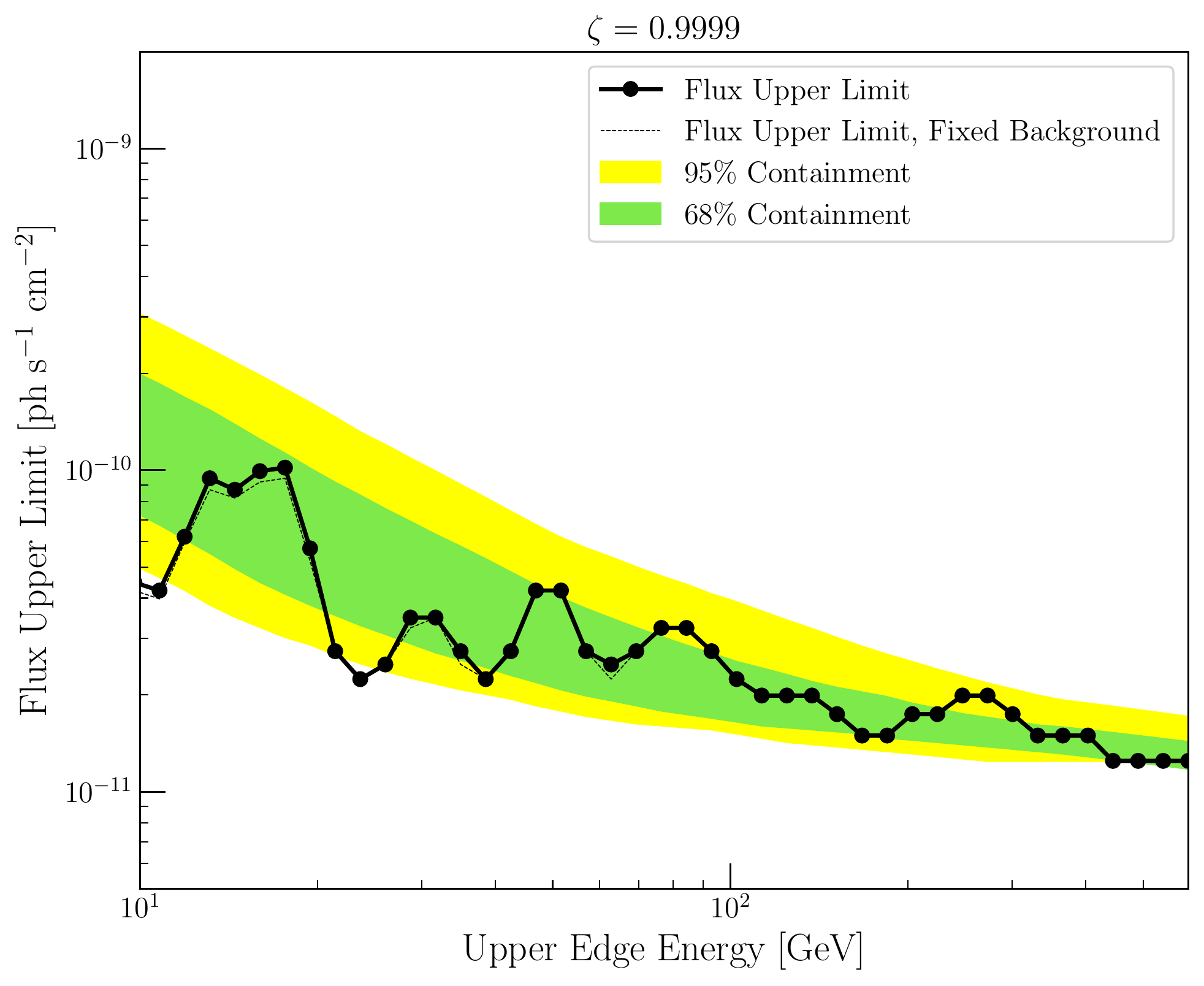}
\noindent
\caption{ 
\label{fig:brazil_lines}
{\it Left Panel:} 95\% confidence flux upper limit on a $\gamma$-ray box point source at the GC with $\zeta = 0.44$. The thin dashed line is the corresponding limit when all background sources are fixed. As expected, fixing the background sources improves the limit at lower energies, though only by a factor of 2 at the most. {\it Right Panel:} The same plot, but for the case of $\zeta = 0.9999$. In both figures, the 68\% and 95\% containment bands come from a Monte Carlo simulation of the data described in Section \ref{sec:MCMethods}. \label{fig:brazil_narrow_box}
}
\end{center}
\end{figure}

No significant signal from a p-wave DM signal was seen in either the case of the wide or narrow box.
The flux upper limits are shown in Figure~\ref{fig:brazil_narrow_box} for both the wide box (left panel) and the narrow box (right panel) scenarios. 
The strongest signal came in the case of $\zeta = 0.44$ at an upper-edge energy 125 GeV; the empirical local significance (found from comparison to the MC TS distribution of Figure \ref{fig:correlations}) was found to be 1.83$\sigma$.
For the case of $\zeta = 0.9999$, the strongest signal came from a box with an upper-edge energy of 84 GeV; the local significance was 1.7$\sigma$.
These do not take into account trials factors, so their global significance is reduced further.

\begin{figure}[ht] 
\begin{center}
\includegraphics[width=0.9\columnwidth]{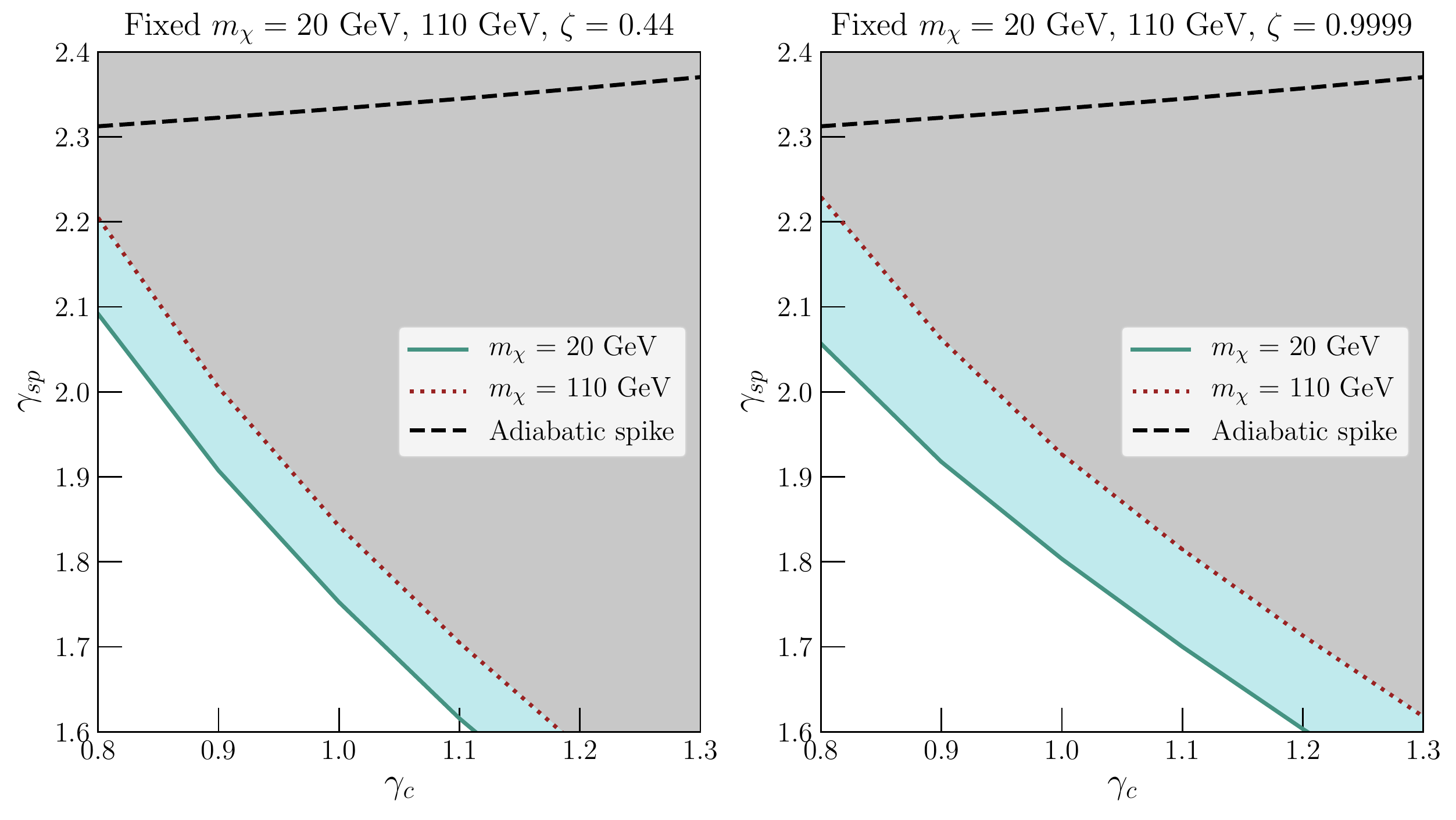}
\noindent
\caption{ 
\label{fig:final_interp}
Shaded regions above each curve show the excluded DM distributions in the GC with thermal relic $p$-wave DM in the hidden sector axion portal model.
Two representative choices of DM mass $m_\chi = 20$ GeV, 110 GeV for narrow boxes (left, $\zeta=0.99$) and wide boxes (right, $\zeta = 0.44$) are shown.
The limits shown here and in Figure \ref{fig:near-final} rely on the assumptions made in Section \ref{sec:dm} about the mass of Sgr A$^*$, the halo velocity dispersion, and the $\gamma \gamma$ branching ratio.
}
\end{center}
\end{figure}

The predicted flux from $p$-wave DM annihilation depends on the DM mass $m_{DM}$ as well as on the power laws of the DM halo ($\gamma_c$) and spike ($\gamma_{sp}$) in our fiducial model.  In Figure~\ref{fig:final_interp} we fix the DM mass, and show how the upper limits on narrow and wide boxes constrain the allowed DM distribution in the GC.  We can observe in particular that adiabatic spikes are excluded for even very shallow cusps $\gamma_c = 0.8$.  In this parameter space, nightmare DM models yielding narrow boxes are less constrained than DM models yielding wide boxes, despite the stronger flux limits; this occurs because the limited phase space available for the narrow box annihilation process further suppresses the annihilation.

\begin{figure}[ht] 
\begin{center}
\includegraphics[width=0.9\columnwidth]{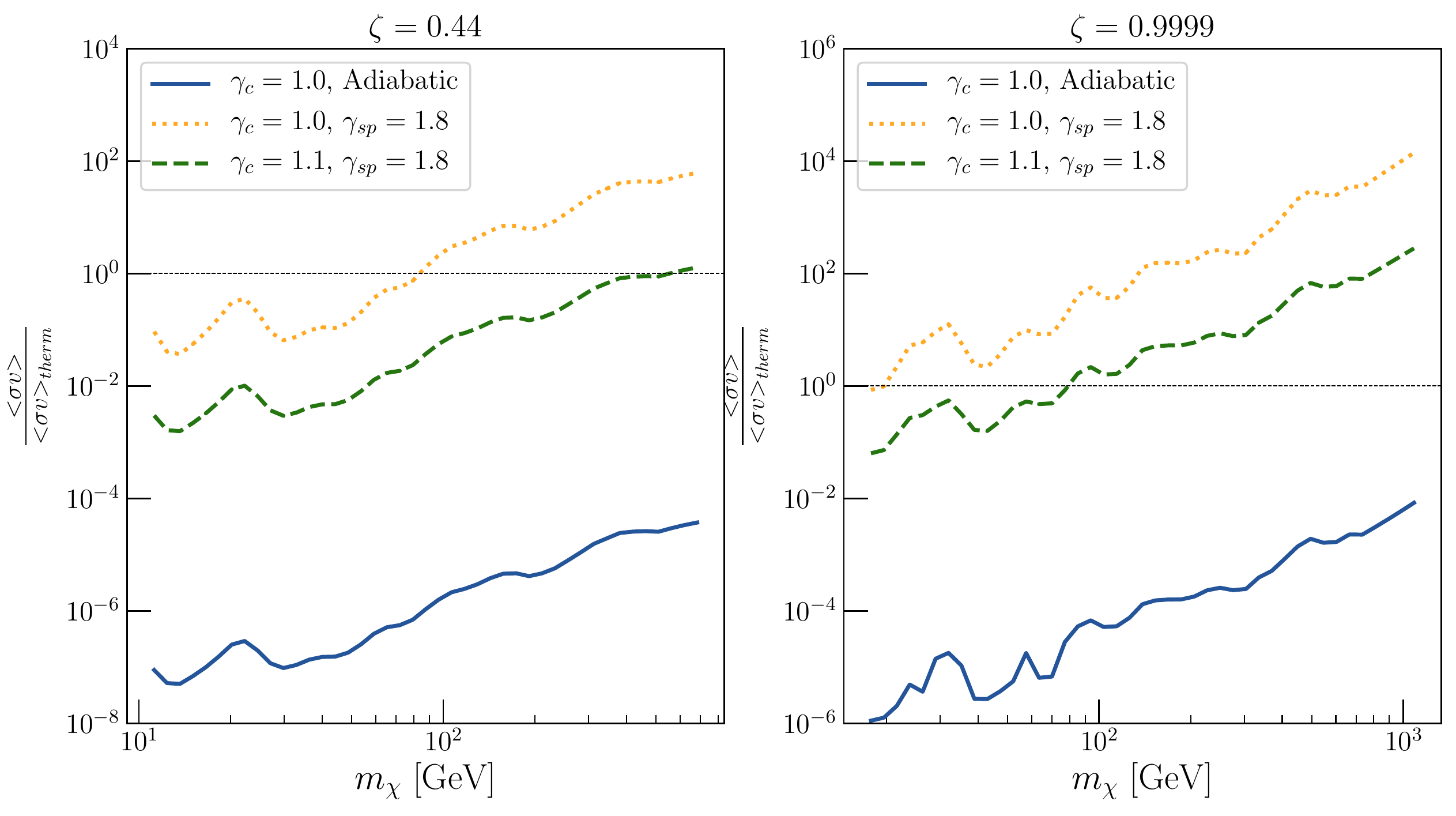}
\noindent
\caption{ 
\label{fig:near-final}
95\% confidence upper limits on DM annihilation cross-section as a function of DM mass for fixed values of $\gamma_c$, $\gamma_{sp}$.
For clarity we plot the velocity-independent ratio of the excluded cross-section to the value that yields the correct relic abundance.  From top to bottom, the three curves correspond to  $\gamma_c = 1.0, \gamma_{sp}=1.8$ (orange);  $\gamma_c = 1.1, \gamma_{sp}=1.8$ (green); and $\gamma_c = 1, \gamma_{sp}=2.33$ (adiabatic spike, blue).
}
\end{center}
\end{figure}

In Figure~\ref{fig:near-final} we consider fixed sample choices of $\gamma_c$ and $\gamma_{sp}$ and show the resulting limits on our reference hidden sector axion portal $p$-wave DM model as a function of DM mass.  For clarity we plot the ratio of the excluded cross-section $\langle\sigma v\rangle$ to the value of the cross-section that yields the correct relic abundance, $\langle\sigma v\rangle_{\mathrm{thermal}}$.  We comment that exclusions for the narrow box scenario in this reference model should not be considered literally at high masses as the model becomes non-perturbative above $m_\chi\sim 300$ GeV.  The need for such large couplings arises to compensate for the phase space suppression that follows when $m_\chi\approx m_\phi$, and no such issue arises in the wide box scenario.

\section{Discussion and Conclusions}\label{sec:conclusion}
In this paper we present the results of a search for the $\gamma$-ray signature of DM annihilating through a $p$-wave channel.
Although most indirect detection searches to date have focused on $\it{s}$-wave annihilation, it is necessary to consider other paradigms in which this channel is suppressed.
As many models of thermal DM have parametrically suppressed couplings to SM particles, and thus no accessible direct detection or collider signals, it
is critical to perform astrophysical searches for such models.
\Fermi-LAT is an ideally suited instrument to perform this search due to its large exposure in the direction of the Galactic center and good energy resolution.
We searched the $\it{Fermi}$-LAT data for the $\gamma$-ray signature of $p$-wave annihilating particle DM at the Galactic center in the energy range 6-800 GeV.
Two spectral models (corresponding to the upper and lower extrema of mediator masses) were tested by comparing the maximum likelihood $\mathcal{L}$ in the presence and absence of a signal.
We found no evidence of a DM signal, and placed an upper limit on the total $\gamma$-ray flux from $\it{p}$-wave annihilation at the center of the Milky Way.

The flux limits presented here are independent of the parameters of the DM spike (i.e. the $J$-factor).
Interpreting these limits further requires making assumptions about the mass of Sgr A$^*$, the halo velocity dispersion, and the branching ratio $Br(\phi\rightarrow\gamma\gamma)$ as described Section \ref{sec:dm}, 
If one assumes a thermal-relic cross section for the annihilation, they can be used to constrain these parameters.
Alternatively, one can use a fixed model of the DM spike to put limits on the annihilation cross section; we found that the annihilation cross section can be constrained to be below the canonical thermal relic cross section given some models of the spike parameters $\gamma_c$ and $\gamma_{sp}$.
Given the two models of mediator masses considered here, it is also possible to use the results in the context of other models of $p$-wave annihilation with intermediate mediator masses. 

\begin{acknowledgments}
The authors would like to thank Bill Atwood for numerous helpful discussions. 

The \textit{Fermi}-LAT Collaboration acknowledges generous ongoing support
from a number of agencies and institutes that have supported both the
development and the operation of the LAT as well as scientific data analysis.
These include the National Aeronautics and Space Administration and the
Department of Energy in the United States, the Commissariat \`a l'Energie Atomique
and the Centre National de la Recherche Scientifique / Institut National de Physique
Nucl\'eaire et de Physique des Particules in France, the Agenzia Spaziale Italiana
and the Istituto Nazionale di Fisica Nucleare in Italy, the Ministry of Education,
Culture, Sports, Science and Technology (MEXT), High Energy Accelerator Research
Organization (KEK) and Japan Aerospace Exploration Agency (JAXA) in Japan, and
the K.~A.~Wallenberg Foundation, the Swedish Research Council and the
Swedish National Space Board in Sweden.
 
Additional support for science analysis during the operations phase is gratefully acknowledged from the Istituto Nazionale 
di Astrofisica in Italy and the Centre National d'\'Etudes Spatiales in France.

Resources supporting this work were provided by the NASA High-End Computing (HEC) Program through the NASA Advanced Supercomputing (NAS) Division at Ames Research Center.

The work of JS was supported by DOE grants DE-SC0015655 and DE-SC0017840.

\end{acknowledgments}

%


\end{document}